\documentclass[prl,showpacs,superscriptaddress,preprint,prl,12pt,onecolumn]{revtex4}%
\usepackage{amsfonts}
\usepackage{amsmath}
\usepackage{amssymb}
\usepackage[dvips]{graphicx}%
\begin{document}

\author{S.~Krompiewski}
\affiliation{Institute of Molecular Physics, Polish Academy of
Sciences, ul.~M.~Smoluchowskiego 17, PL-60179 Pozna{\'n}, Poland}
\author{N.~Nemec}
\affiliation{Institute for Theoretical Physics, University of
Regensburg, D-93040 Regensburg, Germany}
\author{G.~Cuniberti}
\affiliation{Institute for Theoretical Physics, University of
Regensburg, D-93040 Regensburg, Germany}

\title{Spin transport in disordered single-wall carbon nanotubes contacted
to ferromagnetic leads}

\begin{abstract}
Recent conductance measurements on multi-wall carbon nanotubes
(CNTs) reveal an effective behavior similar to disordered
single-wall CNTs. This is due to the fact that electric current
flows essentially through the outermost shell and is strongly
influenced by inhomogeneous electrostatic potential coming from
the inner tubes. Here, we present theoretical studies of
spin-dependent transport through disorder-free double-wall CNTs as
well as single-wall CNTs with Anderson-type disorder. The CNTs are
end-contacted to ferromagnetic electrodes modelled as fcc (111)
surfaces. Our results shed additional light on the giant
magnetoresistance effect in CNTs. Some reported results concern
realistically long CNTs, up to several hundred nanometers.

\end{abstract}

\date{\today}
\maketitle

\section{Introduction}
Over the last two decades, the magneto-electronics, based on
all-metal multilayers, has proven to be very successful indeed
\cite{Prinz}. The most important effect which should be invoked in
this context is giant magnetoresistance (GMR) discovered in 1988
\cite{Baibich}. This effect makes it possible to control electric
current flowing through magnetic materials by means of a magnetic
field. In other words, the essence of the GMR effect lies in
taking advantage of not just the electronic charge alone but also
of its spin counterpart. Quite naturally researchers involved so
far in physics of semiconductors, as well as those studying
molecular systems have intensified their efforts in search for
possible analogous effects in all-semiconducting \cite{Ohno}
and/or hybrid systems (combinations among metals, semiconductors
and molecules) \cite{Tsuka, Xiong}. Consequently, a new field of
science and technology has been triggered, under the name of
spintronics \cite{Zutic, Wolf}. Here we report our results on the
GMR effect in perfect and disordered carbon nanotubes sandwiched
between ferromagnetic electrodes. There is no doubt nowadays that
miniaturization requirements imposed on the emerging spintronics
will be met by applying the so-called bottom-up approach as far as
designing of new electronic devices is concerned. From this point
of view carbon nanotubes are surely excellent candidates.

\section{\textbf{Double-wall CNTs}}

We start our studies with carbon nanotubes (CNTs) end-contacted to
metal electrodes. Our present approach is essentially that
described in detail in \cite{PRB} with an improved simulation
method of CNT/metal-electrode interface developed in
\cite{ustron}. Spin-polarization of the ferromagnetic leads is
defined as
$P=100(n_\uparrow-n_\downarrow)/(n_\uparrow+n_\downarrow)$, where
$n_\sigma$ stands for a number of $\sigma$-spin electrons per
lattice site. It is noteworthy that the structures in question are
relaxed under the Lennard-Jones potential in order to find
energetically favorable relative positions of CNTs' and
electrodes' interface atoms. During the relaxation process the
external electrodes are allowed to rotate and shift independently
of each other, similarly the inner tube is also free to rotate. As
regards the inter-tube hopping integrals, they are taken in the
form as proposed in \cite{roche}, i.e. set to $t_{int} = -(t/8)
{\cos\theta_{lj}} exp[(d_{\,l\,j}-b)/\delta]$,
%
%
where $\theta$ is the angle between the $\pi$ orbitals, d is a
relative distance, $t$ stands for the nearest neighbor hopping
integral (chosen as energy unit), $\delta =$ 0.45 $\AA$ and b =
3.34 $\AA $. The GMR coefficient is defined in terms of the
conductances, $G$, as GMR=$1-G_{\uparrow \downarrow}/G_{\uparrow
\uparrow}$, with $\uparrow \uparrow$ ($\uparrow \downarrow$)
denoting aligned (antialigned) magnetization orientation of the
electrodes.

    Most of the hitherto existing experiments on electronic transport
suggest that current flowing through MWCNTs goes mostly through
the outermost shell (see e.g. \cite{Stojetz}). A precise role of
the inner shells is still hardly known. Here we show the results
on the GMR effect in two double-wall (DW) CNTs which have the same
outer shell but different - though non-conductive in each case -
inner shells. Specifically the DWCNTs in question are: (i) the
zigzag at armchair (45-(5,0)@39-(8,8)) and (ii) the armchair at
armchair (38-(3,3)@ 39-(8,8)), using a short-hand notation
L-(n,m), for the lenghth (in carbon rings) and the chiral vector,
respectively. In the former case the corresponding lengths are
roughly the same (ca. 5 nm each) so both the inner tube and the
outer one are well contacted to the magnetic electrodes. In the
latter case, in turn, the inner shell is artificially shortened
and forced thereby to be out of contact to the drain electrode.
Fig.1 presents giant magnetoresistance for the two DWCNTs. Despite
the fact that both the systems are formally similar (at least in
the presented "energy window" $(\left| E/t \right|<0.2)$, which
falls into the zigzag-tube gap), the GMR curves are clearly
different. We attribute these differences to intertube-quantum
interferences which are present owing to the non-vanishing
$t_{int}$. To mimic a possible effect of some additional disorder
we present also GMR curves (thin lines) calculated from the
energy-averaged conductances, where the averaging has been made
over the most obvious energy scale in this context, namely over an
energy bin equal to the inter-level spacing of the outer shell
$\Delta E = \pi \sqrt{3}/L $ (in the present units). In the
following subsection we present a more direct approach to the
disorder issue.

\section{\textbf{Single-wall CNTs with Anderson disorder}}

The simplest way to include the effect of disorder in a system
described in terms of the tight-binding model is to allow all
on-site (atomic) potentials to take random values within a given
energy interval. Such an approach has been already applied to CNTs
\cite{Triozon}, but to our knowledge, only for tubes with
non-magnetic leads. In the case of disordered systems, it is
necessary to perform statistical (ensemble) averaging of the
results corresponding to particular sets of on-site potential
distributions (to be referred to as samples hereafter). This is a
purely technical problem easy to overcome at the expense of the
computation time. Another more serious problem is to develop a
recursive procedure, which would make it possible to deal with big
systems approaching macroscopic sizes of the order of several
hundred nanometers. Recursive algorithms based on Dyson-type
equations for the Green's function are well-known \cite{Lake,
KMB}. Here however we modify those methods in order to make them
work in the case of highly non-homogenous systems composed of
disordered carbon nanotubes and two adjacent atomic layers from
each electrode (to be referred to as the extended molecule). While
conductance computations are usually rather fast, the computations
of electronic charge at all atoms of a big system are extremely
computer time consuming and expensive. In order to surmount this
difficulty we impose a global charge neutrality condition only on
a disorder-free "parent" system and self-consistently determine
detailed values of all its on-site potentials (ca. 4000 and 40000
atoms for the SWCNT(8,8), 30 and 300 nm long). On introducing
disorder, these on-site potentials are modified by random
corrections fluctuating around zero within an interval [-W/2,
W/2]. So, on the average the global charge of the
Anderson-disordered extended molecule might be regarded as roughly
close to that of the neutral parent system. Our computations
proceed according to the following protocol: First the surface
Green functions are found (see \cite{ustron} for details). Second,
the set of on-site potentials which ensure the charge neutrality
of the parent system is found. Third, conductance calculations
along with the corresponding GMR coefficients are performed for
100 different samples with random on-site corrections. Finally the
results are ensemble-averaged. The main results of this study is
presented in Fig.2, for the SWCNT (8,8) consisting of 240 carbon
rings (120 unit cells $\sim$ 30 nm). It is seen that although
disorder suppresses the GMR, it happens to be of about the right
value
as compared to recent experiments on MWCNTs with transparent
ferromagnetic contacts made from $Pd_{0.3}Ni_{0.7}$ (device
resistances are then as low as 5.6 $k\Omega$ at 300K)
\cite{Sahoo}. Other noteworthy points are: (i) on the average the
GMR remains positive, and (ii) there exist some extra features in
the GMR spectrum at energies close to $\pm0.4$ and $\pm0.7$
corresponding to higher sub-band onsets in the pristine (ideal)
SWCNT.

For smaller $W$, GMR increases and eventually oscillates with the
amplitude of roughly $\pm 0.2$ in the disorder-free (parent) case,
as shown in Fig.3 (l.h.s). Additionally the right-hand side of
this figure highlights the length-effect on the period of
oscillations. For the sake of simplicity this is shown for the
paramagnetic leads. It is clearly seen that in the absence of
disorder the observed periodicity reflects length-dependent
interferences, as expected for a (quasi) ballistic transport. In
the magnetic case the peaks are split due to the lifting of
spin-degeneracy. The quasi-periodic behavior does always occur
when there is no disorder, regardless of whether the electrodes
are ferromagnetic or not (c.f. the inset in Fig.3 with the thick
solid curve on the r.h.s). From the present results one sees that
the process of averaging of conductance and GMR spectra leads to a
subtle interplay between the length and the amount of disorder in
the CNTs.

\section{Summary}

In this work it has been shown theoretically that the GMR effect
in ferromagnetically contacted carbon nanotubes is quite
considerable and may reach a few tens percent. Ideally, the GMR
coefficient oscillates as a function of energy (gate-voltage) with
a quasi-period close to the inter-level spacing of the CNT, which
scales inversely proportional to the nanotube length. Yet, such a
picture is to some extent too detailed if the system at hand is
imperfect, e.g. due to some impurities, dopants or a presence of
incommensurate inner shells in a MWCNT. The disorder-averaged GMR
rages from 6\% down to 2\% in the vicinity of the charge
neutrality point,  in conformity with recent experiments on MWCNTs
with transparent ferromagnetic contacts. Furthermore, the
aforementioned periodicity gets nearly completely suppressed, and
there is no more tendency for the GMR to become negative.

\vspace{1cm}

\begin{acknowledgments}
 S. K. thanks the KBN project
(PBZ-KBN-044/P03-2001), the Centre of Excellence (contract No.
G5MA-CT-2002-04049) and the Poznan Supercomputing and Networking
Center for the computing time. This work was partially funded by
the Volkswagen Foundation and the Vielberth Foundation.
\end{acknowledgments}

\newpage

\begin{figure}[ht]
\includegraphics[width=.5\textwidth]{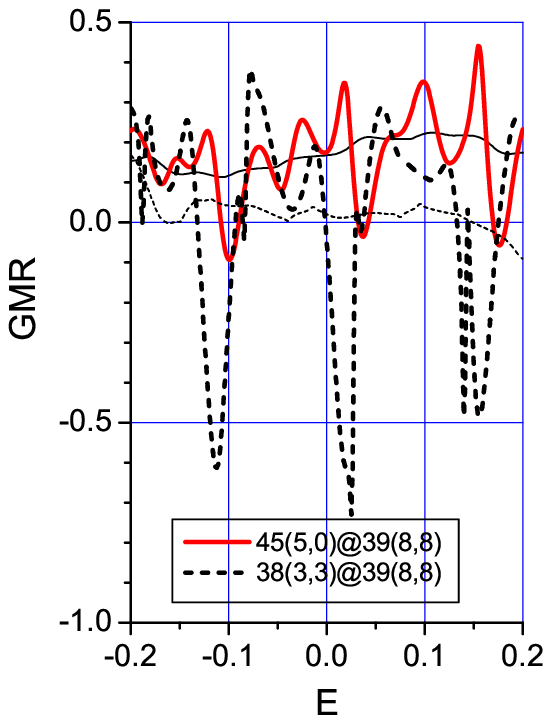}
 \caption{GMR for the double-wall carbon nanotubes
45-(5,0)@39-(8,8) (solid thick line) and 38-(3,3)\@39-(8,8) (thick
dashed line) attached to ferromagnetic leads with 50\%
spin-polarization. To mimic a possible effect of disorder, there
are also shown the GMR curves computed from $\Delta E$-averaged
conductances (thin curves of the same style), where $\Delta E$ is
the inter-level spacing of the outer shell.} \label{fig1}
\end{figure}

\newpage

 \begin{figure}[htb]
 \includegraphics[width=.45\textwidth]{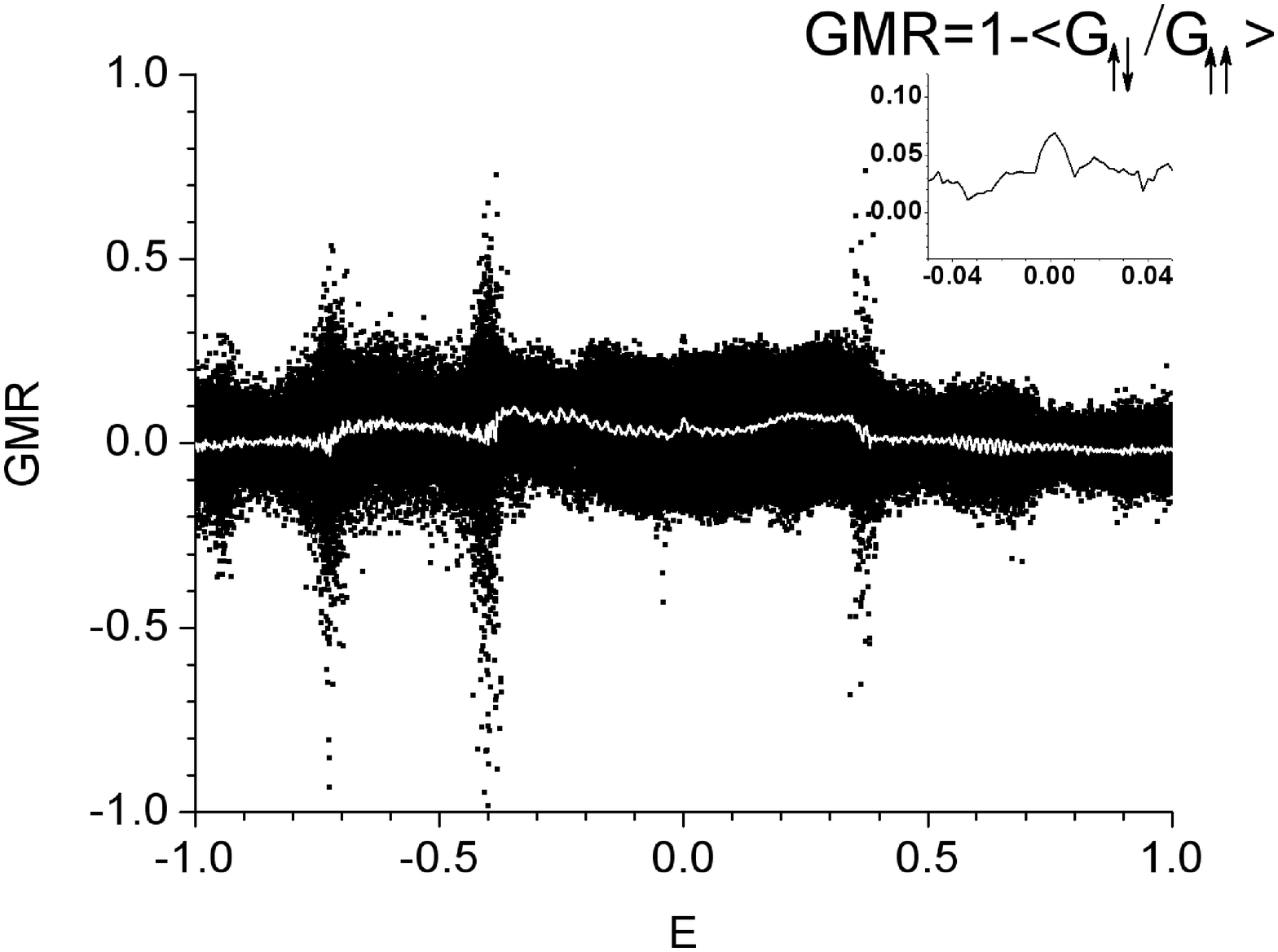}
 \hfil
\includegraphics[width=.45\textwidth]{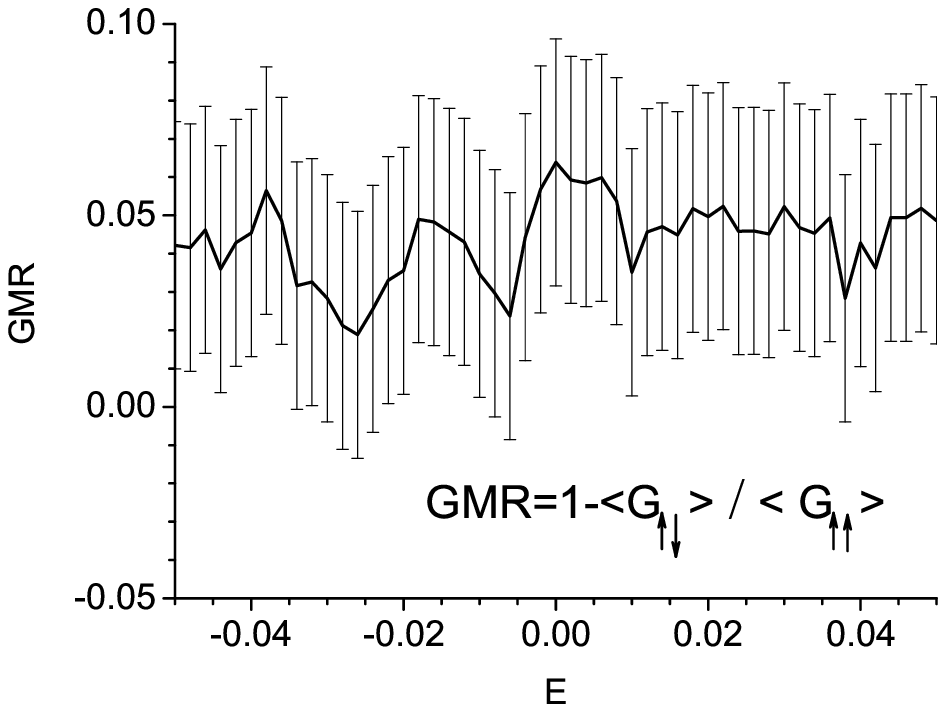}
\caption{Left hand side: GMR for individual SWCNTs (8,8), ca. 30
nm in length (points), along with the GMR (white curve) averaged
over 100 samples with disorder-induced corrections to the on-site
potentials (within [-W/2, W/2] for W=0.2). On the right hand site
the GMR computed from the disorder-averaged conductances together
with the standard-deviation error bars are shown in the vicinity
of the charge neutrality point.} \label{fig:12}
\end{figure}

\newpage
 \begin{figure}[htb]
 \includegraphics[width=.45\textwidth]{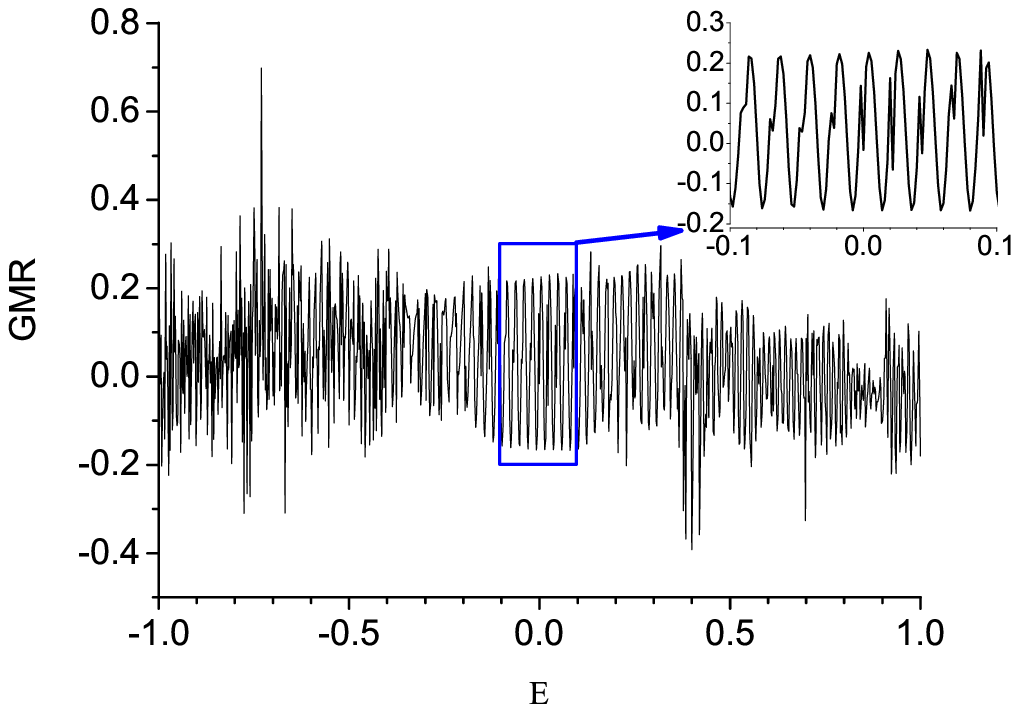}
 \hfil
 \includegraphics[width=.45\textwidth]{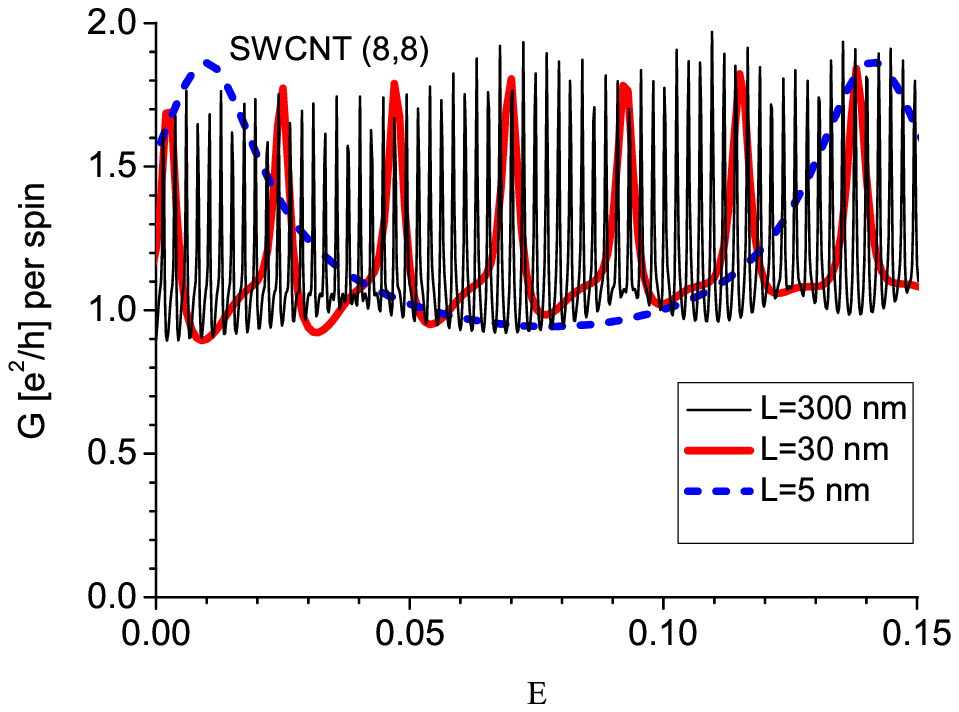}
\caption{Left hand side: GMR for disorder-free SWCNT (8,8),
P=50\%, W=0, L=240 carbon rings ($\sim$ 30nm). Right hand side:
visualization of the length-dependent periodicity of the
conductance for the case of paramagnetic leads, P=0, and L=5, 30
and 300 nm. Compare the inset with the thick solid curve to see
that the quasi-period of oscillations is roughly maintained, but
in the magnetic case the peaks are spin-split.} \label{fig:3}
\end{figure}

\end{document}